\documentclass[12pt]{article}

\usepackage{amsmath,amssymb}
\usepackage{amsfonts}
\usepackage{graphicx}
\usepackage{a4wide}

\usepackage[T2A]{fontenc}
\usepackage[cp1251]{inputenc}
\usepackage[russian]{babel}

\begin{document}

\begin{center} {\bf M. V. Burnashev} \end{center}

\begin{center}
{\large\bf ON THE EXACT DECODIND ERROR PROBABILITY EXPONENT OF THE RANDOM CODING ON BSC}
\footnote[1]{The research is supported by  MSHE RF GZ project}
\end{center}

{\begin{quotation} \normalsize For the information transmission a
binary symmetric channel is used. The transmission of
exponential number of messages is 
considered. The exact decoding error probability exponent 
is derived. The proof is based on the new results on the distribution 
of a certain sum of random variables. 

\emph{Key words and phrases:} random coding, decoding error probability exponent. 
\end{quotation}}

\begin{center}
I. INTRODUCTION AND MAIN RESULTS
\end{center}

Consider the binary symmetric channel {\rm BSC}$(p)$ with crossover 
probability $p$, $0 \leq p \leq 1/2$ and $q = 1-p$ \cite{Elias1, F,G1, VO}.
Consider also a binary code ${\mathcal C_n}(R)$ of block length 
$n$ with $M = e^{Rn}$, $R \geq 0$, codewords $\boldsymbol{x}_1,\ldots, \boldsymbol{x}_M$. 
Each codeword $\boldsymbol{x}_i$, $i=1,\ldots,M$ consists of $n$ binary input symbols, 
i.e., $\boldsymbol{x}_i = (x_{i,1},\ldots,x_{i,n})$, where $x_{i,j} \in \{0,1\}$ 
for each $j$, $1 \leq j \leq n$. Moreover, it is assumed that 
all $x_{i,j}$ are i.i.d random variables, taking equiprobable values $0$ and $1$.  

Denote by $\boldsymbol{y}\in \{0,1\}^n$ the channel $n$-vector output on the final 
instant $n$ and denote
\begin{equation}\label{deftau1}
\begin{gathered}
d_{i} =d(\boldsymbol{y},\boldsymbol{x}_{i}), \quad i=1,\ldots,M, \quad
k=1,\ldots,n,
\end{gathered}
\end{equation}
where $d(\boldsymbol{y},\boldsymbol{x})$ is the Hamming distance between 
$\boldsymbol{y}$ and $\boldsymbol{x}$.

At the final instant $n$ we make the decision $\hat{\boldsymbol{x}}$
in favor of the message $\boldsymbol{x}_{i}$ with the 
smallest $d_{i}$, and denote by $P_{\rm e}({\mathcal C_n},p,R)$ the decoding error probability
for a particular code ${\mathcal C_n}$
\begin{equation}\label{defPe}
\begin{gathered}
P_{\rm e}({\mathcal C_n},p,R) = \max_{\boldsymbol{x}_{i} \in \mathcal C_n}
{\mathbf P}\{\hat{\boldsymbol{x}} \neq \boldsymbol{x}_{i}|\boldsymbol{x}_{i}\}
\end{gathered}
\end{equation}
We choose a code ${\mathcal C_n} = \{\boldsymbol{x}_1,\ldots, \boldsymbol{x}_M\}$ 
randomly,  such that each codeword $\boldsymbol{x}_i$ consists of $n$ independent 
random variables, taking equiprobable values $0$ and $1$.  


Denote by $P_{\rm e}(n,p,R) = {\mathbf E}_{{\mathcal C_n}}P_{\rm e}({\mathcal C_n},p,R)$ the 
expected value (over all codes ${\mathcal C_n}$) of the decoding error probability 
$P_{\rm e}({\mathcal C_n},p,R)$ for such randomly chosen codes ${\mathcal C_n}$.
Define the random coding exponent $E_{r}(R)$ as
\begin{equation}\label{defErR}
\begin{gathered}
E_{r}(R) = \lim_{n \to \infty}\frac{-\ln P_{\rm e}(n,p,R)}{n}. 
\end{gathered}
\end{equation}

In addition to the traditional critical rate $R_{\rm cr}(p)$ \cite{G1, G2, VO}
\begin{equation} \label{mainres3}
\begin{gathered}
R_{\rm cr}(p) = \ln 2 - h(b_{0}), \quad b_0 = \frac{\sqrt{p}}{\sqrt{q}+\sqrt{p}}, 
\quad 0 \leq p \leq 1/2, 
\end{gathered}
\end{equation}
introduce also the new critical rate $R_{\rm crit}(p)$ 
\begin{equation} \label{defRcr}
\begin{gathered}
R_{\rm crit}(p) = \frac{(\sqrt{q}-\sqrt{p})}{2(\sqrt{q}+\sqrt{p})}\ln(q/p), 
\quad 0 \leq p \leq 1/2. 
\end{gathered}
\end{equation}
Then the inequality holds 
\begin{equation} \label{mainres3a}
\begin{gathered}
R_{\rm crit}(p) < R_{\rm cr}(p), \quad 0 < p < 1/2. 
\end{gathered}
\end{equation}

In order to simplify formulas we represent the rate $R$, 
$0 \leq R \leq C(p) = \ln 2 -h(p)$ as 
\begin{equation} \label{defR}
\begin{gathered}
R = \ln 2 -h(\delta_{R}), \quad 0 \leq p \leq \delta_{R} \leq 1/2, \\
h(x) = -x\ln x-(1-x)\ln(1-x), \quad 0 \leq x \leq 1.
\end{gathered}
\end{equation}

The main result of the paper represents

{\it Theorem 1}. For the random coding exponent $E_{r}(R)$ the following formula holds 
\begin{equation} \label{mainres1}
\begin{gathered}
E_{r}(R) = g(p,R), \\
\end{gathered}
\end{equation}
where 
\begin{equation} \label{mainres1a}
\begin{gathered}
g(p,R) = \left\{\begin{array}{ccc}
2R - \ln 2 + 2h(r_{0}) + \ln(\sqrt{pq}), &0 \leq R \leq R_{\rm crit}(p); \\
                  \ln 2 - 2\ln\left(\sqrt{q}+\sqrt{p}\right)- R, &R_{\rm crit}(p) \leq R \leq R_{\rm cr}(p); \\
                  E_{\rm sp}(R),
                   &
                  R_{\rm cr}(p) \leq R \leq C(p),  
                \end{array}
\right. \\
\end{gathered}
\end{equation}
and
\begin{equation} \label{mainres2}
\begin{gathered}
r_{0} = \frac{1}{2} + \frac{R}{\ln(p/q)}, \quad C(p) = \ln 2- h(p), \\
E_{\rm sp}(p,R) = \delta_{R}\ln\frac{\delta_{R}}{p} + (1-\delta_{R})\ln\frac{1-\delta_{R}}{q}.  
\end{gathered}
\end{equation}

{\it Remarks}. 
1) Similar to \eqref{mainres1}-\eqref{mainres1a} relations are well known
\cite[formula (5.6.45)]{G1}, \cite[formula (3.4.5)]{VO} for 
$R_{\rm crit}(p) \leq R \leq  C(p)$,  
but, in fact, all of them have the form of lower bounds 
 for $E_{r}(R)$, (i.e. $E_{r}(R) \geq g(p,R)$. 
 On the other hand, it is known that for small $R$ there exist  codes 
 (expurgation codes) for which the coding exponent
 is greater than the right-hand side of \eqref{mainres1}. 
 The question that arises is whether the weakness of $E_{r}(R)$
 at low rates $R$ is due to the bounding technique used in \cite{G1}
 and that question was investigated in  \cite{G3}.
In a sense, this paper supplements the paper \cite{G3}, 
giving the exact asymptotic of $E_{r}(R)$ for all rates $R$. 

2) R. Gallager wrote in \cite[after formula (5.6.45)]{G1}
``The most significant point about this example is that even for such a simple channel 
there is no simple way to express $E_{\rm r}(R)$ except in parametric form''. 
Nevertheless, if we except the representation $R = \ln 2 -h(\delta)$ as natural, 
then formulas \eqref{mainres1} - \eqref{mainres2} for $E_{\rm r}(R)$ are not 
parametric ! 

\begin{center}
II. PROOFS
\end{center}
 
Note that ${\mathbf p}(\boldsymbol{y}|\boldsymbol{x}) = q^{n}
z^{d(\boldsymbol{y},\boldsymbol{x})}$, $z = p/q$. 
Denote by $\pi_{i}$, $i=1,\ldots,M$, the posterior probability of $\boldsymbol{x}_{i}$ 
at the instant $n$. Then  
\begin{equation} \label{mainres3}
\begin{gathered}
\pi_{i} = \frac{z^{d_{i}}}{S(z,M,n)}, \quad 
S(z,M,n) = \sum\limits_{j=1}^{M}z^{d_{j}}, \quad i=1,\ldots,M, 
\quad z = \frac{p}{q}.
\end{gathered}
\end{equation}
Also, if $\boldsymbol{x}_{\rm true} = \boldsymbol{x}_{m}$, then
\begin{equation} \label{mainres3a}
\begin{gathered}
\pi_{m} = \frac{z^{d_{m}}}{z^{d_{m}} + T(z,M,n)}, \qquad
T(z,M,n) = \sum\limits_{j \neq m}z^{d_{j}}. 
\end{gathered}
\end{equation}
First, we shall consider the distribution of the random sum $S(z,M,n)$ from \eqref{mainres3},
and then use that distribution to investigate the posterior probability 
$\pi_{\rm true} = \pi_{m}$.
Note that $d_{j} = d(\boldsymbol{y},\boldsymbol{x}_{j}) = 
w(\boldsymbol{y}\bigoplus \boldsymbol{x}_{j})$, where $w(\boldsymbol{z})$ means the weight 
(i.e. the number of ones) of binary $n$-vector $\boldsymbol{z}$, and 
$\boldsymbol{y}\bigoplus \boldsymbol{x}$ means the sum by modulo $2$. Since the 
codewords $\{\boldsymbol{x}_{j}\}$ are independently chosen, the distribution of the 
sum $S(z,M,n)$ does not depend on $\boldsymbol{y}$, and we may set 
$\boldsymbol{y}= \boldsymbol{0}$. Then $d_{j} = w(\boldsymbol{x}_{j})$ and when 
defining $S(z,M,n)$  we may set in \eqref{mainres3} 
$z^{d_{j}} = z^{w(\boldsymbol{x}_{j})}$ for all $j$.

{\it A. Auxiliary result}. 

Consider a randomly chosen binary code of block length $n$ with $M = e^{Rn}$, 
$R > 0$, codewords $\boldsymbol{x}_1,\ldots, \boldsymbol{x}_M$. 
Denote by $w(\boldsymbol{x}_{j})$, $j=1,\ldots,M$ - weight 
of binary $n$-vector $\{\boldsymbol{x}_{j}\}$. Consider the random sum
\begin{equation} \label{sum2}
\begin{gathered}
S(z,M,n) = \sum\limits_{j=1}^{M}z^{w_{j}(n)}, \qquad z > 0.
\end{gathered}
\end{equation}
The following result was proved in \cite{Bur23}.

{\it Theorem 2}.
For any $z > 0$ and $n \geq 10$ the following inequality holds
\begin{equation} \label{Cor1a}
\begin{gathered}
{\mathbf P}\left\{\left|\ln\frac{S(z,M,n)}{Mz^{n/2}}\right| \geq
\sqrt{n\ln(n+1)}|\ln z|\right\} \leq (n+1)^{-M}.
\end{gathered}
\end{equation}

{\it Remark}. It follows from \eqref{Cor1a} that if $M \sim e^{Rn}$, $R > 0$,
then $S(z,M,n) \sim Mz^{n/2}$ for any $z > 0$ with very high probability.


{\it B. Proof of Theorem 1}.

Let $\boldsymbol{x}_{\rm true} = \boldsymbol{x}_{m}$. 
The posterior probability $\pi_{m}$ of $\boldsymbol{x}_{m}$ at instant $n$ is
described by \eqref{mainres3a}. At instant $n$ we make the decision in 
favor of the message $i$ with the largest $\pi_{i}$ (i.e. with the smallest 
$d_{i}$). Therefore, the decoding error may happen only if $\pi_{m} < 1/2$
(since all remaining probabilities $\pi_{i}$ are positive). We denote 
$d_{i} = b_{i}n$, $i=1,2\ldots,M$ and $d_{m}= b_{m}n$.  Then we get
by \eqref{mainres3a} for the decoding error probability $P_{r}(n)$ of 
random coding 
\begin{equation} \label{dec3a1}
\begin{gathered}
P_{r}(n) = {\mathbf E}_{d_{m}}{\mathbf P}\{\min_{i \neq m} d_{i}\leq d_{m}|d_{m}\} =
{\mathbf E}_{d_{m}}{\mathbf P}\{\text{error};\pi_{m} < 1/2|d_{m}\} = \\
= {\mathbf E}_{d_{m}}{\mathbf P}\{\text{error}; T(z,M,n) > z^{d_{m}}|d_{m}\} 
\end{gathered}
\end{equation}
and the upper bound (it will be used for $R \geq R_{\rm crit}$)
\begin{equation} \label{dec3a2}
\begin{gathered}
P_{r}(n) \leq {\mathbf E}_{d_{m}}{\mathbf P}\{T(z,M,n) > z^{d_{m}}|d_{m}\}. 
\end{gathered}
\end{equation}


For random coding the values $T = T(z,M,n) = S(z,M-1,n)$ and $d_{m}$ are independent 
random variables with known distributions (see \eqref{Cor1a}). 
We are interested in the probability of the event $\{d_{i} \leq d_{m}\}$ for one 
(or more) indices $i \neq m$. We have
\begin{equation} \label{Auxil1}
\begin{gathered}
{\mathbf P}\{\text{error}|d_{m},T\} = 
{\mathbf P}\{\min_{i \neq m} d_{i} \leq d_{m}|d_{m},T\}, 
\end{gathered}
\end{equation}
and by \eqref{Cor1a} 
\begin{equation} \label{Cor1a1}
\begin{gathered}
{\mathbf P}\left\{\left|\ln\frac{T(z,M,n)}{(M-1)z^{n/2}}\right| \geq
\sqrt{n\ln(n+1)}|\ln z|\right\} \leq (n+1)^{-(M-1)}.
\end{gathered}
\end{equation}
Denote (see also \eqref{defRcr})
\begin{equation} \label{defrho}
\begin{gathered}
r_{0} = \frac{1}{2} + \frac{R}{\ln z}, \\ 
r = \frac{1}{2} - \sqrt{\frac{\ln (n+1)}{n}}  + \frac{\ln(M-1)}{n\ln z} = r_{0} + 
O\left(\sqrt{\frac{\ln n}{n}}\right). 
\end{gathered}
\end{equation}
Note that the value $rn$ is the minimal distance $d_{m}$ for which the decoding 
error may happen. Then we have by \eqref{mainres3a} and \eqref{Cor1a1}-\eqref{defrho}
\begin{equation} \label{Auxil3a}
\begin{gathered}
{\mathbf P}\{\pi_{m} < 1/2|d_{m}\} = 
{\mathbf P}\{S(z,M-1,n) > z^{d_{m}}|d_{m}\} = \\
={\mathbf P}\left\{
(M-1)z^{n/2-\sqrt{n\ln(n+1)}} >z^{d_{m}}|d_{m}\right\} + O(n^{-M}) = 
I_{\{d_{m} > r n\}} + O(n^{-M}), 
\end{gathered}
\end{equation}
and by \eqref{dec3a1} and \eqref{Auxil3a}
\begin{equation} \label{Auxil3}
\begin{gathered}
P_{r}(n) = 
{\mathbf E}_{d_{m}}{\mathbf P}\{\text{error}; S(z,M-1,n) > z^{d_{m}}|d_{m}\} = \\
= {\mathbf E}_{d_{m}}{\mathbf P}\left\{\text{error};d_{m} > r n|d_{m}\right\} + O(n^{-M})
= {\mathbf E}_{\{d_{m}> r n\}}{\mathbf P}\left\{\text{error}|d_{m}\right\} + O(n^{-M}), 
\end{gathered}
\end{equation}
where ${\mathbf E}_{\cal A}\xi = {\mathbf E}\left(\xi I_{\cal A}\right)$. 
We also have for any $i \neq m$ and $d_{m} =b_{m}n$, $b_{m} \leq 1/2$ 
\begin{equation} \label{dec3a1a2}
\begin{gathered}
\frac{1}{n}\ln P\{d_{i}\leq d_{m}\} = 
\frac{1}{n}\ln \left[2^{-n}\sum_{i=0}^{b_{m}n}\binom{n}{i}\right] = 
h(b_m) - \ln 2 + o(1), \quad n \to \infty.
\end{gathered}
\end{equation}
Then
\begin{equation} \label{dec3a1a}
\begin{gathered}
\frac{1}{n}\ln {\mathbf P}\left\{\text{error}|d_{m}\right\} =
\frac{1}{n}\ln\left\{MP\{d_{i}\leq d_{m}\} \right\} + o(1) = \\
= - [\ln 2 -R-h(b_m)]_{+} + o(1). 
\end{gathered}
\end{equation}
The relation \eqref{dec3a1a} is sufficiently accurate, since, for example,
$$
\begin{gathered}
\frac{1}{2}MP\{d_{i}\leq d_{m}\} \leq {\mathbf P}\left\{\text{error}|d_{m}\right\} \leq  
MP\{d_{i}\leq d_{m}\},  \quad \text{if} \  MP\{d_{i}\leq d_{m}\} \leq \frac{1}{2}. 
\end{gathered}
$$

For the true distance $d_{m}$ we have $P\{d_{m}\} = q^{n}\binom{n}{d_{m}}z^{d_{m}}$ and then
\begin{equation} \label{newed1}
\begin{gathered}
\frac{1}{n}\ln {\mathbf P}\left\{d_{m} = bn\right\} = f_{1}(p,b)+ o(1), \ n \to \infty, \\
f_{1}(p,b) = \ln q + h\left(b\right) + b\ln z, \quad 0 \leq b \leq  1. 
\end{gathered}
\end{equation}
The function $f_{1}(p,b)$ is concave on $b \geq 0$. Also $f_{1}(p,p) = 0$ and 
$f_{1}(p,b) < 0$ for $b \neq p$.

Then using the representation \eqref{defR} we have 
\begin{equation} \label{newed2}
\begin{gathered}
\frac{1}{n}\ln{\mathbf P}\left\{d_{m} \geq \delta_{R} n\right\} = 
f_{1}(p,\delta_{R}) + o(1) = -E_{\rm sp}(p,R) + o(1).
\end{gathered}
\end{equation}

Thus, we get by \eqref{Auxil3} and \eqref{dec3a1a}-\eqref{newed1}
with $d_{m}=b_{m}n$
\begin{equation} \label{newed3}
\begin{gathered}
P_{r}(n) = \left[1 + O(n^{-M})\right]
{\mathbf E}_{\{b_{m} \geq r\}}{\mathbf P}\left\{\text{error}|b_{m}n\right\}, 
\end{gathered}
\end{equation}
and
\begin{equation} \label{newed4}
\begin{gathered}
\frac{1}{n}\ln P_{r}(n) = \frac{1}{n}\ln\left[
{\mathbf E}_{\{b_{m} \geq r\}}{\mathbf P}\left\{\text{error}|b_{m}n\right\}\right] + 
O(n^{-(M+1)})  = \\
= \max_{b \geq r}\left\{- [\ln 2 -R-h(b)]_{+} + 
f_{1}(p,b)\right\} + O(n^{-(M+1)}) = \\
= \max_{b \geq r_{0}}f_{2}(p,R,b) + o(1),
\end{gathered}
\end{equation}
where 
\begin{equation} \label{newed4ab}
\begin{gathered}
f_{2}(p,R,b) = - [\ln 2 -R-h(b)]_{+} + \ln q + h\left(b\right) + b\ln z = \\
= - [h(\delta_{R}) - h(b)]_{+} + \ln q + h\left(b\right) + b\ln z, \qquad 
r_{0} = \frac{1}{2} + \frac{R}{\ln z}, 
\end{gathered}
\end{equation}
and $r, r_{0}$ are defined in \eqref{defrho}. Also,  $\delta_{R} \geq p$. 

To perform maximization in \eqref{newed4} consider two possible 
cases depending on 
$[h(\delta_{R}) - h(b)]_{+}$: \\
$1)\ \delta_{R} \geq \max\{p,b\}$, and $2)\ p \leq \delta_{R} < b$. 

Case 
$1)\ \delta_{R} \geq \max\{p,b\}$. Then 
\begin{equation} \label{newed4abc}
\begin{gathered}
f_{2}(p,R,b) = R-\ln 2 + \ln q + 2h\left(b\right) + b\ln z. 
\end{gathered}
\end{equation}
Note that 
\begin{equation} \label{newed4b}
\begin{gathered}
(f_{2})'_{b} = 2\ln\frac{1-b}{b} + \ln z, \quad (f_{2})''_{bb} < 0, 
\quad (f_{2})'_{b=0}= + \infty.  
\end{gathered}
\end{equation}
Therefore, the function $f_{2}(p,R,b)$ from \eqref{newed4abc} 
is concave on  $b$ and it has the unconditional 
maximum on $b$ when $(f_{2})'_{b} = 0$, i.e for $b = b_0$ (see \eqref{mainres3}).

In \eqref{newed4} we need $\delta_{R} \geq \max\{p,b\}$ and
$\delta_{R} \geq b \geq r_{0}$.  Since the function $f_{2}(p,R,b)$ has its 
unconditional  maximum at $b = b_0$, there are possible two subcases: 
$1a) \ b_0 \leq r_{0}$, and $1b) \ b_0 \geq r_{0}$.

Note that by \eqref{defrho}, the condition $b_0 \leq r_{0}$ is equivalent to have
\begin{equation} \label{newed5}
\begin{gathered}
R \leq \frac{(\sqrt{p}-\sqrt{q})}{2(\sqrt{q}+\sqrt{p})}\ln z = R_{\rm crit}(p). 
\end{gathered}
\end{equation}
On the contrary, the condition $b_0 \geq r_{0}$ is equivalent to have
\begin{equation} \label{newed6ab}
\begin{gathered}
R \geq \frac{(\sqrt{p}-\sqrt{q})}{2(\sqrt{q}+\sqrt{p})}\ln z = R_{\rm crit}(p). 
\end{gathered}
\end{equation}

Consider cases $1a)$ and $1b)$. 

Case $1a) \ b_{0} \leq r_{0}$ and $\delta_{R} \geq b \geq r_{0}$. 
Note that the following inequality holds
\begin{equation} \label{newed4a1}
\begin{gathered}
r_{0} \leq \delta_{R}, \quad 0 \leq R \leq 1.
\end{gathered}
\end{equation}
Then maximum in \eqref{newed4abc} is attained for $b = r_{0}$ and 
\eqref{newed4abc} takes the form 
\begin{equation} \label{newed4a}
\begin{gathered}
\frac{1}{n}\ln P_{r}(n) = f_{2}(p,R,r_{0}) + o(1), 
\quad R \leq R_{\rm crit}(p), \\
f_{2}(p,R,r_{0}) = 2R - \ln 2 + 2h(r_{0}) + \ln(\sqrt{ pq}).    
\end{gathered}
\end{equation}

Case $1b) \ b_{0} \geq r_{0}$ and $\delta_{R} \geq b \geq r_{0}$. 
Then maximum in \eqref{newed4abc} is attained 
for $b = b_{0}$, what gives 
\begin{equation} \label{newed6}
\begin{gathered}
\frac{1}{n}\ln P_{r}(n) = f_{2}(p,R,b_{0}) + o(1), \\
h(b_0) = h\left(\frac{\sqrt{p}}{\sqrt{q}+\sqrt{p}}\right) = 
\ln\left(\sqrt{q}+\sqrt{p}\right) - 
\frac{\sqrt{p}\ln p + \sqrt{q}\ln q}{2(\sqrt{q}+\sqrt{p})}, \\
f_{2}(p,R,b_0) = R - \ln 2 + \ln q+ 2h\left(b_0\right) +b_0\ln z =  \\
= R - \ln 2 + 2\ln\left(\sqrt{q}+\sqrt{p}\right), \quad 
R_{\rm crit}(p) \leq R \leq R_{\rm cr}(p).  
\end{gathered}
\end{equation}
We need also $\delta_{R} \geq b_{0}$, i.e. $h(\delta_{R}) \geq h(b_{0})$, what means
(see \eqref{mainres3})
\begin{equation} \label{newed61}
\begin{gathered}
h(\delta_{R}) = \ln 2 - R \geq h(b_{0}) = 
\ln\left(\sqrt{q}+\sqrt{p}\right) - 
\frac{\sqrt{p}\ln p + \sqrt{q}\ln q}{2(\sqrt{q}+\sqrt{p})},  \\ 
R \leq \ln 2 - h(b_{0}) = R_{\rm cr}(p), \quad 
b_0 = \frac{\sqrt{p}}{\sqrt{q}+\sqrt{p}}, \quad 0 \leq p \leq 1/2, .
\end{gathered}
\end{equation} 

Case $2)\ \delta_{R} < b$. Then \eqref{newed4} takes the form 
\begin{equation} \label{newed41}
\begin{gathered}
\frac{1}{n}\ln P_{r}(n) = 
\max_{b \geq \max\{r_{0},\delta_{R}\}}f_{1}(p,b) + o(1), \\
f_{1}(p,b) = \ln q + h\left(b\right) + b\ln z.
\end{gathered}
\end{equation}
The function $f_{1}(p,b)$ is concave on $b \geq 0$ and 
it has the unconditional maximum $f_{1}(p,p) = 0$
for $b=p$. Also, $\delta_{R} \geq p$ and 
$(f_{1})'_{b} < 0$ for $b > p$. By \eqref{newed4a1} the maximum in 
\eqref{newed41} is attained for $b = \delta_{R}$ and we have (see \eqref{newed2})
\begin{equation} \label{newed41a}
\begin{gathered}
\frac{1}{n}\ln P_{r}(n) = 
\max_{b \geq \max\{r_{0},\delta_{R}\}}
f_{1}(p,b) + o(1) = f_{1}(p,\delta_{R}) + o(1) = -E_{\rm sp}(p,R) + o(1).
\end{gathered}
\end{equation}

From \eqref{newed4a}--\eqref{newed41a}  
Theorem 1 with \eqref{mainres1} - \eqref{mainres2} follow.


Marat V. Burnashev 

Higher School of Modern Mathematics MIPT,  1 Klimentovskiy per., Moscow, Russia \\

{\it marat.burnashev@mail.ru}


\begin{center} {\large REFERENCES} \end{center}
\begin{enumerate}
\bibitem{Elias1}
{\it Elias P.} Coding for Noisy Channels. 
IRE Convention Record Part 4, pp. 37-46. 1955. 
\bibitem{F}
{\it Fano R. M.} Transmission of Information. MIT Press, 
Cambridge, Mass and   Wiley, New York, 1961.
 
\bibitem{G1}
{\it Gallager R. G.} Information theory and reliable communication.
Wiley, NY, 1968.
\bibitem{G2}
{\it Gallager R. G.} A Simple Derivation of the Coding Theorem and
some Applications // IEEE Trans. Inform. Theory. 1965. V. 11.
P. 3--18.
\bibitem{G3}
{\it Gallager R. G.} The Random Coding Bound Is Tight for the Average Code 
// IEEE Trans. Inform. Theory. 1973. V. 19.
P. 244--246.
\bibitem{VO}
{\it Viterbi A. J., Omura J.K.} Principles of digital communication and coding.
McGraw-Hill, New York, 1961. 
\bibitem{Bur23}
{\it Burnashev M. V.} On the Distribution of a Statistical Sum 
    Related to the Binary Symmetric Channel // Problems of  
    Information Transmission.  V. 61, no. 1, pp. 11–7, 2025.
    https://doi.org/10.1134/S0032946025010016
\end{enumerate}
\end{document}